\documentclass[]{article} 
\usepackage{graphicx} 
\usepackage[margin=1in]{geometry}
\usepackage{amsmath, amsfonts, amssymb}
\usepackage{color}
\usepackage{bm}
\usepackage{array}
\usepackage{comment}
\usepackage{amsthm}
\usepackage{float}
\usepackage{lipsum}
\usepackage{url}
\usepackage{cite}
\DeclareGraphicsExtensions{.pdf,.png,.jpg}
\usepackage{times}
\usepackage[
  draft=false,
  colorlinks=true,
  linkcolor=red,
  citecolor=red,
  urlcolor=red
]{hyperref}
\usepackage[hang]{caption}
\usepackage[subrefformat=parens]{subcaption}
\title{\Large{Democratization of Real-time Multi-Spectral Photoacoustic Imaging: Open-Sourced System Architecture for\\OPOTEK Phocus \& Verasonics Vantage Combination}}

\author{Ryo Murakami$^{1}$, Yichuan Tang$^{1}$, and Haichong K. Zhang$^{*,1,2,3}$
\\
\footnotesize $^1$Robotics Engineering, $^2$Biomedical Engineering, $^3$Computer
Science, Worcester Polytechnic Institute, MA, United States\\
\footnotesize $^*$Corresponding author: Haichong K. Zhang, hzhang10@wpi.edu,\\
\footnotesize $^*$ORCiD: Ryo Murakami: 0009-0005-5770-5866, Haichong K. Zhang:
0000-0002-1314-8456,\\
\footnotesize $^*$Keywords: Real-time Imaging, Photoacoustic Imaging, System Architecture, Shared Memory, Open-Source.}
\date{}

\begin{document}
\maketitle

\begin{abstract}
Real-time multi-spectral photoacoustic imaging (RT-mPAI) often suffers from synchronization instabilities when interfacing fast-tuning lasers with data acquisition platforms executing on non-real-time operating systems. To overcome this, we establish an open-source hardware-software architecture tailored for the widely adopted combination of the OPOTEK Phocus lasers and Verasonics Vantage systems. By employing an independent micro-controller for deterministic laser trigger counting alongside a decoupled client-server data streaming framework, the proposed system circumvents OS-induced timing deviations and local storage bottlenecks.
By open-sourcing this pipeline and cultivating a collaborative environment to share both code and ideas, we aim to lower the technical and cost barriers for RT-mPAI, thereby democratizing access to stable RT-mPAI research and, more ambitiously, fostering a vibrant open-source community.
\end{abstract}

\begin{figure}[H]
    \centering
    \includegraphics[width=1.0\textwidth]{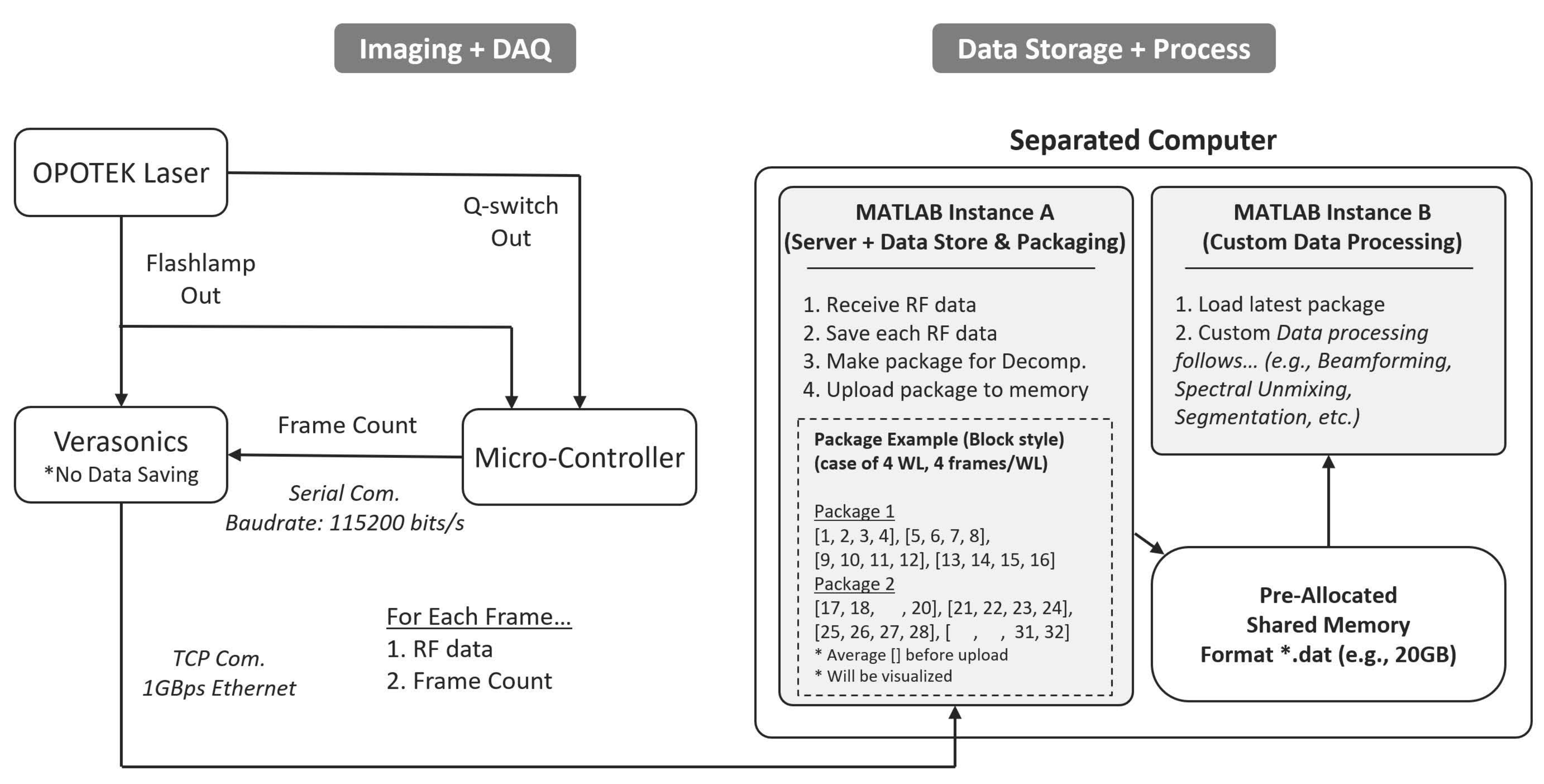}
    \caption{Overall architecture of the proposed RT-mPAI system.
    The hardware layer incorporates an independent micro-controller for deterministic laser trigger monitoring, while the software layer utilizes a decoupled client-server framework to stream data via shared memory.
    (Please note that the "Data Storage + Process" part is optional for the minimal implementation of RT-mPAI. The "Imaging + DAQ" part, with saving and processing on Verasonics, should be enough.)}
    \label{fig:arch}
\end{figure}

\section{Introduction}
\label{sec_intro}
Real-time multi-spectral photoacoustic imaging (RT-mPAI) is a promising candidate for real-time functional imaging \cite{taruttis2015advances,oeri2017hybrid,jeng2021real,murakami2023closed}. 
Intraoperative guidance is considered a major application where RT-mPAI can demonstrate its high potential \cite{lediju2020photoacoustic,wiacek2021photoacoustic,bell2026listening}.
Examples include visualizing blood vessels and nerves during laparoscopic surgery to enhance procedural safety \cite{gao2026handbook}, as well as monitoring ablation therapies \cite{basij2022integrated,gao2021necrosis,gao2026handbook,murakami2024intraoperative,murakami2024preliminary,murakami2024thermal,murakami2026interstitial,murakami2026direct}.
To realize RT-mPAI, it is essential to correctly assign the laser wavelength as metadata to each image frame, and this must be implemented in a high-speed, reliable manner.
The objective of this report is the democratization of RT-mPAI: lowering the barrier to entry for RT-mPAI and increasing the number of environments capable of performing RT-mPAI, even with existing system configurations.
While systems capable of readily achieving RT-mPAI certainly exist, they are often difficult to adopt due to factors such as cost.

The combination of the OPOTEK Phocus laser series and the Vantage system from Verasonics is a widely adopted setup within the PAI community \cite{sankepalle2025custom}.
The primary appeal of this combination lies in the exceptional flexibility it offers for both wide-range wavelength tuning and ultrasound data acquisition \cite{kratkiewicz2021technical,chen2022transparent}.
Especially, the fast-tuning feature of the OPOTEK Phocus series is highly attractive for mPAI, as users can define a sequence of wavelengths to be fired.

However, a major challenge when utilizing the fast-tuning feature is that the programmed wavelength sequence is fixed once transmitted to the laser, preventing proactive wavelength adjustments during acquisition. Furthermore, the control software for the OPOTEK Phocus series currently lacks an interface to query real-time wavelength information.
Given that precise wavelength tagging of individual laser pulses is a strict prerequisite for downstream spectral analysis, resolving this informational ambiguity is critical.
As Sankepalle \textit{et al.} \cite{sankepalle2025custom} pointed out, a viable approach is to monitor the hardware triggers from the laser and deterministically assign a wavelength to each frame. This strategy is effective because the OPOTEK Phocus system reliably adheres to the pre-programmed sequence.
To make this approach work, the minimum but critical requirement is that the Verasonics system needs to capture all the triggers from the laser (e.g., 20 Hz) without missing any. 
For this minimum requirement, the Verasonics Vantage system has one inherent challenge: the host computer is running on the non-real-time Windows operating system (Microsoft Corporation, Redmond, WA, United States), and the Verasonics program needs to communicate with the MATLAB (MathWorks, Inc., Natick, MA, United States) software running on Windows.
This challenge in real-time operation can affect the trigger monitoring stability, especially when the image size is large, when the saving function is used, or when the host computer itself is busy (e.g., due to other processes running on Windows).
Consequently, from a system reliability standpoint, the timing instabilities introduced by the non-real-time environment must be treated as an inevitable occurrence rather than a hypothetical risk, especially in the context of RT-mPAI.

Given these technical challenges, we propose the introduction of an independent external micro-controller, which is dedicated to monitoring and counting the triggers from the laser, and providing the counter to the Verasonics system.
With this architecture, even if the Verasonics data acquisition cannot maintain its desired repetition frequency, the micro-controller can keep counting, and the following data processing for RT-mPAI can correctly assign a wavelength to the acquired image.

Furthermore, as a supplementary feature, we have introduced structural enhancements to facilitate seamless RT-mPAI workflows. Specifically, we propose a configuration that bypasses local data storage on the Verasonics host. Instead, the acquired RF data, paired with its corresponding hardware frame counter, is transmitted frame-by-frame to an independent receiving computer via TCP communication. By eliminating local storage overhead, this architecture theoretically enables stable, temporally unlimited continuous data acquisition. Moreover, because the receiving computational environment can be freely tailored by the user to optimize post-processing, this approach affords complete flexibility in choosing the operating system and programming languages. Consequently, this decoupled architecture significantly lowers the barrier for integrating GPU acceleration and utilizing widely distributed AI tools, particularly those within the Python ecosystem.

Therefore, the expected major benefits can be summarized as follows:

\begin{enumerate}
    \item \textit{\textbf{Higher accessibility to RT-mPAI under the combination of the OPOTEK Phocus laser and Verasonics Vantage system}}
    \item \textit{\textbf{Virtually unlimited \& stable scanning}}
    \item \textbf{\textit{High flexibility in real-time data processing (e.g., GPU, programming language, and operating system)}}
\end{enumerate}

Lastly, although this approach represents a highly specific solution tailored for the combination of OPOTEK Phocus and Verasonics Vantage systems, we believe there is substantial merit in thoroughly documenting and disseminating this concept, given the widespread adoption of this hardware pairing within the community. Therefore, we have decided to open-source the core code components and system architecture diagrams required to implement this synchronization framework.

While the current release provides an open-source environment optimized for this specific hardware configuration, our ultimate objective is to foster a collaborative ecosystem where diverse use cases, improvements, and broader insights regarding RT-mPAI can be shared, transcending hardware brand limitations.
We hope this initiative will establish a foundational platform that democratizes access to and utilization of RT-mPAI for the entire research community. Furthermore, we respectfully hope that the technical insights derived from this proposal might serve as a modest reference for future commercial product developments. Ultimately, it is our ideal vision that continued advancements in native system integrations will eventually render such external synchronization workarounds entirely unnecessary.

\section{Materials \& Methods}
\subsection{System Overview}
\label{sec_overview}
The proposed system mainly consists of two components: 1) imaging \& data acquisition, and 2) data storage \* processing (Figure \ref{fig:arch}).
The core concept of this report is implemented in the first component, where both the Flashlamp out and Q-switch out triggers are monitored by a micro-controller, and the Verasonics system is triggered by the Flashlamp out trigger.
The acquired image data with the corresponding frame number is transferred to a separate computer for data storage and processing.


\subsection{Components 1: Imaging \& Data Acquisition}
\label{sec_image_daq}
\begin{figure}[H]
    \centering
    \includegraphics[width=0.8\textwidth]{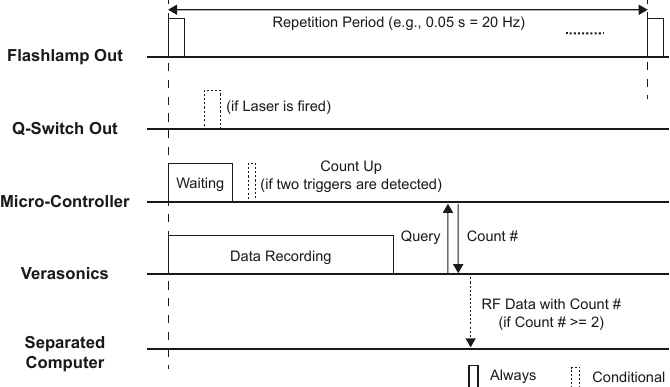}
    \caption{Timing and sequence diagram of the entire imaging system.
    The micro-controller deterministically counts laser pulses by monitoring the Flashlamp out and Q-switch out triggers.
    The Verasonics system queries the count number and transmits it with the image data if the image data is categorized as effective.}
    \label{fig:timing}
\end{figure}

The primary concept of this proposal is implemented in this component. The key is that the micro-controller monitors the laser triggers with sufficiently high frequency and high reliability. In this example, the Arduino platform is used. 
The OPOTEK Phocus laser provides a flashlamp out trigger first during the preparation phase, and provides both flashlamp out and Q-switch out when the effective laser is fired.
Therefore, the micro-controller is programmed to count up the frame counter only when both triggers are detected.
Although theoretically only monitoring the Q-switch trigger should be sufficient, it also monitors the flashlamp out trigger to enhance the detection stability and warm up the following TCP communication (Figure \ref{fig:timing}).
The channel data recording by the Verasonics system is triggered by the flashlamp out signal.
Once it finishes the recording for the corresponding laser pulse, the Verasonics system communicates with the micro-controller to inquire the current frame number. 
If the count number is 2 or more, the Verasonics system packs the channel data and count number into a single packet and transfers it through TCP communication.
Here, the initial count is intentionally ignored because the fast-tuning feature of the OPOTEK Phocus series is designed to fire a dummy pulse and start the programmed wavelength sequence.

\subsection{Components 2: Data Storage \& Process}
\label{sec_store_process}
In this example, two MATLAB instances are established on another computer connected with the Verasonics host computer through TCP communication. A pre-allocated memory is accessible by both instances for a fast data transfer within the computer. Please note that Component 2 can be implemented by using other environments, such as C++ and Python, depending on the objectives. Extensions to these two languages are at least already considered for future development.

\subsubsection{Server Side: Data Reception and Pre-processing}
The server instance is dedicated to receiving data through TCP communication and to packaging the RF data based on the programmed wavelength sequence for the following client side.

\begin{figure}[H]
     \centering
     \includegraphics[width=0.8\textwidth]{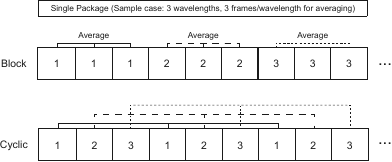}
     \caption{Schematics of data packaging performed by the server instance.
     Both block and cyclic scanning schemes are supported.}
     \label{fig_package}
 \end{figure}

The server supports two data packaging schemes: \textit{block} (sequential frames per wavelength) and \textit{cyclic} (interleaved frames) layouts, structuring the averaged data into a standardized tensor (e.g., $\text{Axial} \times \text{Lateral} \times \text{Wavelengths}$). This tensor is then immediately written to a shared memory block (Figure \ref{fig_package}), and the raw single frame data before averaging is stored on the server side.

\subsubsection{Client Side: Custom Data Processing}
The client side operates independently of the server's reception loop.
The client continuously polls the shared memory for newly processed packages.
This instance executes the core, application-specific data processing.
Since the averaging and wavelength sorting are already done, the client side is not required to perform these pre-processing steps and can directly move on to specific tasks, including beam-forming \cite{treeby2010k,treeby2012modeling,treeby2010modeling}.
Due to the decoupled nature of the system, computationally heavy tasks, such as spectral unmixing/decomposition, external hardware triggering, can be executed without causing buffer overflows on the reception end.
Moreover, as already mentioned above, this architecture gives users significant flexibility in selecting program languages, GPUs, and libraries, etc.

\subsection{System Validation}
\label{sec_validation}
To validate the proposed pipeline's real-time wavelength identification capability, a phantom experiment is performed.
The standard scan mode of the OPOTEK Phocus control software is used as a ground-truth because the wavelength being fired is shown on the screen, making it easy to assign a wavelength to each frame after the entire data collection is completed.
As a test group, the proposed real-time wavelength identification is performed in real-time using the fast-tuning mode.
The test group frames are classified into each wavelength based on the wavelength assignment map generated in real-time.

\subsubsection{Phantom Design}
\begin{figure}[H]
    \centering
    \includegraphics[width=0.6\textwidth]{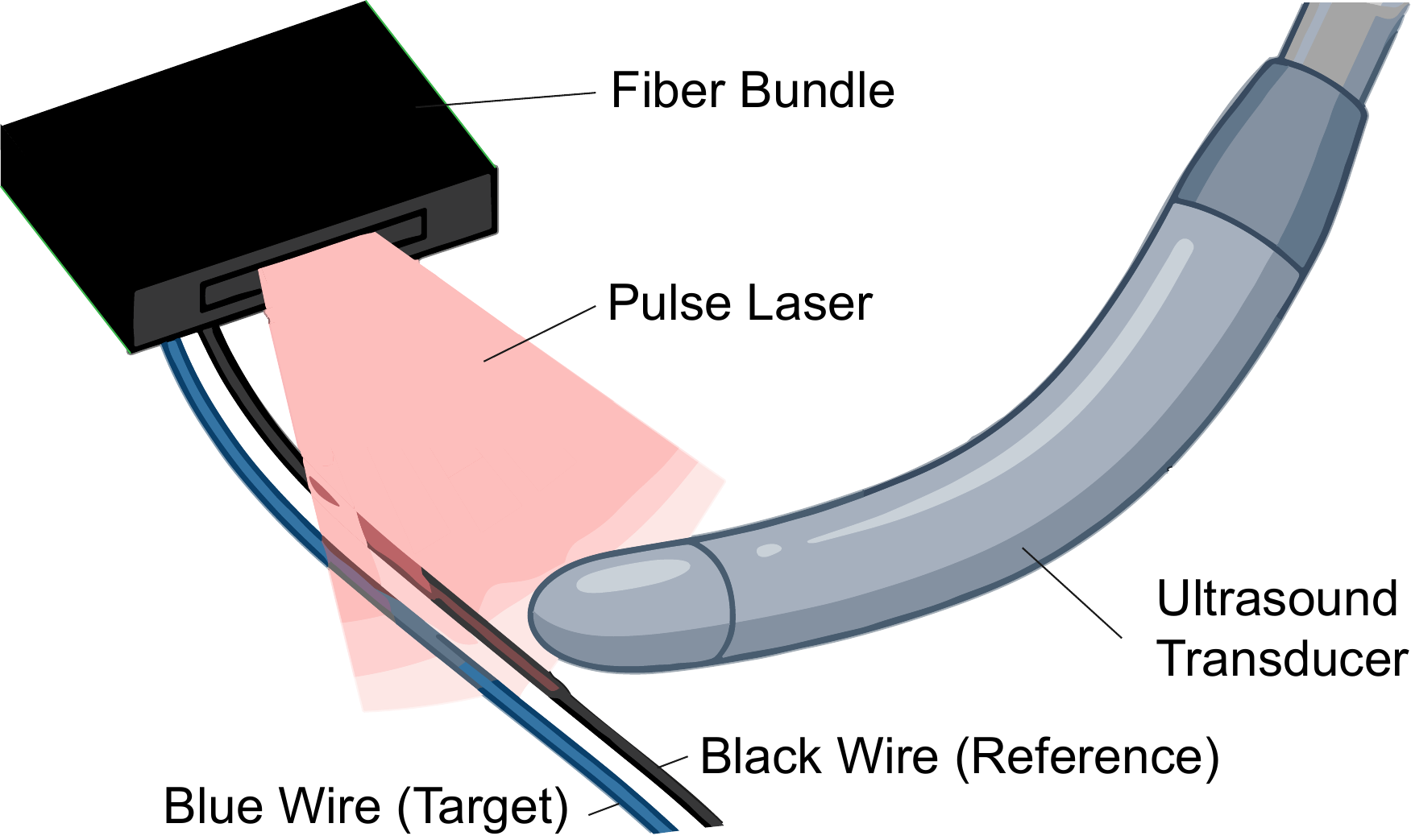}
    \caption{Schematic of the validation setup.
    Blue and black wires are utilized as point targets.
    Calculating the signal ratio between the two wires effectively compensates for incidental pulse-to-pulse laser energy fluctuations.}
    \label{fig:phantom_design}
\end{figure}

Blue and black wires are used as point targets representing different spectra.
They are fixed inside the water, and mPAI is performed, including both wires (Figure \ref{fig:phantom_design}).
To compensate for the laser energy fluctuation, the ratio of the PA signal from the blue wire with respect to the one from the black wire is computed for each frame as a primary PA signal.

\subsubsection{Imaging \& Data Acquisition Setup}
The experimental setup for this validation utilizes a PAI system comprising an OPOTEK Phocus mobile laser (Phocus MOBILE, OPOTEK, Inc., Carlsbad, CA, United States) operating at a $20$ Hz repetition rate, and a Verasonics Vantage 128 platform (Verasonics, Kirkland, WA, United States) equipped with a Philips ATL Lap L9-5 ultrasound transducer (Philips, Amsterdam, Netherlands).
An Arduino micro-controller (Arduino MEGA, Somerville, MA, United States) serves as the hardware synchronization unit.
Laser pulses are delivered via a fiber bundle at wavelengths of $700$, $740$, $760$, and $780$ nm.
For both standard and fast-tuning (cyclic) modes, $500$ frames are acquired and averaged per wavelength.

\subsubsection{Results}
\begin{figure}[H]
    \centering
    \includegraphics[width=0.8\textwidth]{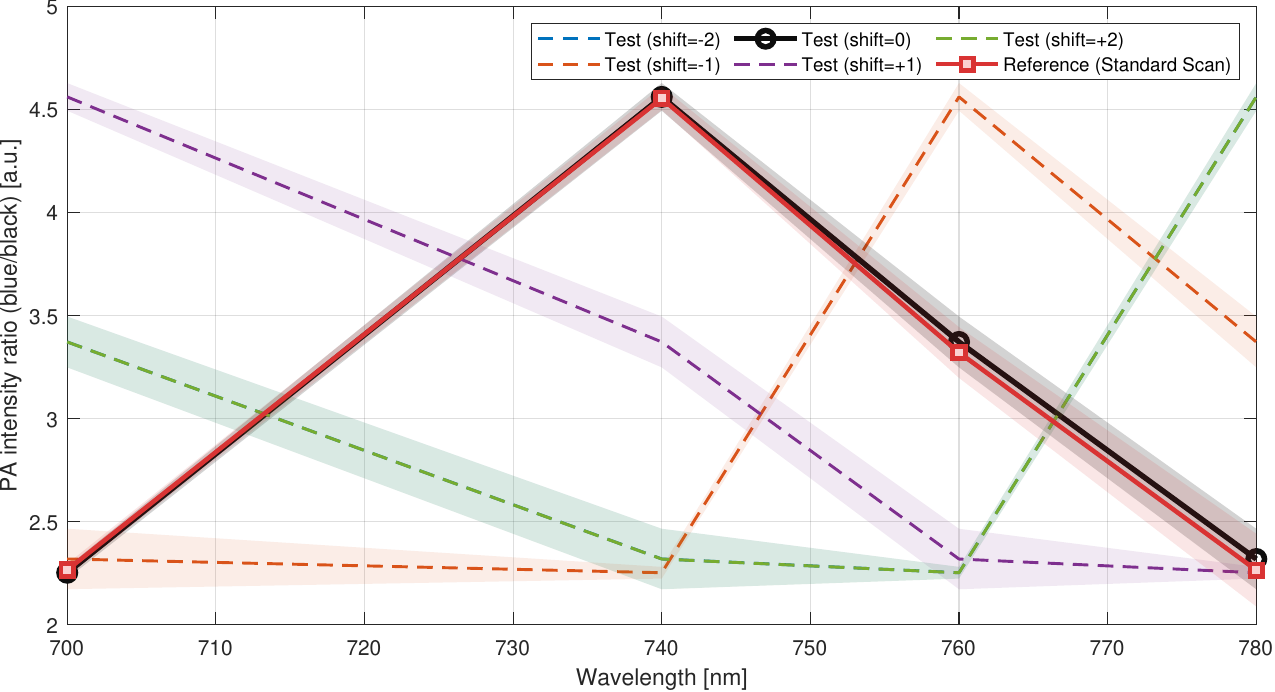}
    \caption{Validation of the real-time wavelength identification pipeline. The spectrum acquired via the proposed system in fast-tuning mode (black) correctly aligns with the ground-truth standard scan reference (red). Introducing artificial frame shifts ($\pm1$, $\pm2$) to the real-time dataset yields clearly diverted spectra (blue, orange, purple, green), demonstrating the necessity and robustness of the hardware trigger counting method. NOTE: The data of $\text{shift}=-2$ is overlapped by that of $\text{shift}=+2$.}
    \label{fig_spec}
\end{figure}

As shown in Figure \ref{fig_spec}, the spectrum acquired via the proposed real-time wavelength identification pipeline using the fast-tuning mode (black line) demonstrates agreement with the reference spectrum obtained from the ground-truth standard scan (red line).
This close correspondence highlights the accuracy of the wavelength assignment and supports the validity of the proposed pipeline.
To further validate the synchronization logic, artificial frame shifts of $\pm1$ and $\pm2$ were introduced to the real-time dataset.
These shifts assume potential trigger counting errors to evaluate whether an alternative wavelength sequence could yield a more accurate spectrum. The resulting shifted spectra (depicted in blue, orange, purple, and green in Figure \ref{fig_spec}) exhibit deviations from the reference spectrum, unlike the aligned zero-shift case (black).
The clear degradation in spectral accuracy under these simulated desynchronization conditions additionally confirms the validity and robustness of the real-time trigger counting pipeline.
In total, the same scans were repeated three times and five times for the reference and test case, respectively. The same trend is observed in the other cases. Although a slight intensity difference is observed, we concluded that it can be attributed to the wavelength fluctuation.

\section{Discussion}
\label{sec_discussion}
In this report, we propose adding one external micro-controller for independent and stable laser trigger monitoring to enable a reliable RT-mPAI with the combination of the OPOTEK Phocus laser and Verasonics Vantage system.
As far as we have tested, we have not observed any miscounting on the Arduino side.
Although we sometimes observed missed frames, the independent counting by Arduino enabled the server MATLAB instance to treat the missed frames properly.
For this particular setup, the server instance ignores the missing frames in the averaging.
This data handling would slightly degrade the image quality for that specific package, but the package after averaging still can keep the expected structure (e.g., $\text{Axial} \times \text{Lateral} \times \text{Wavelength}$) unless the pipeline misses all the frames for a wavelength.

\subsection{Future Development \& Limitations}
\begin{enumerate}
    \item \textbf{\textit{Real-time pulse-to-pulse energy compensation:}}
    
    For an ideal RT-mPAI, the laser energy compensation needs to be performed pulse-to-pulse in real-time.
    The real-time interrogation of the energy measurement is now being integrated into the proposed pipeline.
   	This feature is expected to be implemented in the updated version of our code repository.
   	At this moment, we measure the laser energy for each wavelength in advance and use these fixed values for RT-mPAI.
    
    \item \textbf{\textit{Laser delay compensation:}}
    
    The current setup assumes the Verasonics system starts the data acquisition with a constant delay after receiving the Flashlamp out signal.
    However, the delay slightly varies. We believe this is due to the synchronization between the Verasonics system and its host MATLAB environment.
    This variation causes a minor axial shift in channel data.
    Although this effect does not significantly impact the current results, it is a parameter that fundamentally requires precise control.
    A potential improvement currently under consideration involves sending a trigger to the micro-controller from the Verasonics system once its recording is ready.
    This signal would be detected by the micro-controller to measure the time difference relative to the Flashlamp out signal, and this temporal information would then be transmitted via TCP alongside the RF data and frame counter. While routing the Verasonics trigger directly to the laser's Q-switch in port is a commonly employed method, we intentionally avoid this configuration to ensure robust RT-mPAI operation. Specifically, we have observed that when the trigger from the Verasonics system is delayed due to the timing instabilities, the laser fails to trigger appropriately, resulting in the omission of the corresponding laser pulse.

    \item \textbf{\textit{Compatibility with other probes:}}
    
    The current pipeline is only tested and optimized for the specific probe (Philips ATL Lap L9-5 ultrasound transducer).
    Generalization of the pipeline to incorporate other probes is needed.

    \item \textbf{\textit{C++ pipeline development for accelerated performance:}}
     
    As mentioned above, the other end of the TCP communication can be any computation environment as long as it can receive and process data appropriately.
    The development of the C++ pipeline is on our roadmap, and it is expected to be available on our GitHub repository once it's ready.
    
    \item \textbf{\textit{Further validation with spectrometer:}}
    
    In the current validation, we focused on evaluating whether the acquired spectral curve aligns with the reference, as this serves as a practically important metric.
    Furthermore, considering the hardware specifications, wavelength synchronization should logically be correct at the individual pulse level.
    However, for a more rigorous future validation, it is necessary to verify the emitted wavelength on a pulse-by-pulse basis using a spectrometer.

\end{enumerate}

\section{Conclusion}
\label{sec_conclusion}
In this report, we introduced an open-source hardware-software architecture to enable reliable RT-mPAI using OPOTEK Phocus lasers and Verasonics Vantage systems.
By integrating an independent micro-controller for deterministic trigger monitoring and a decoupled client-server data streaming framework, the proposed system successfully overcomes inherent timing instabilities.
Through this open-source initiative, we aim to lower technical barriers, thereby democratizing access to stable RT-mPAI and fostering further collaborative advancements within the research community.

\section*{Code Availability}
\label{sec_code}
The core components of the primary features are available online.
We actively welcome feedback, suggestions for improvements, and bug reports from the community (Contact regarding code: Ryo Murakami, \href{mailto:rmurakami@wpi.edu}{rmurakami@wpi.edu}).\\(URL: \url{https://github.com/RyoMurakami-MedRobo/Realtime-Multiwavelength-PA-Toolkit}).


\section*{Acknowledgement}
\label{sec_acknowledge}
The authors gratefully acknowledge Shun Katayose, M.D., for his valuable contribution to the setup and organization of the GitHub page.
This work is supported by the National Institutes of Health under grants (CA134675, OD028162).

\section*{Disclaimer}
\label{sec_disclaimer}
The contents, system architectures, and observations described in this report are based on the authors' independent research and experience utilizing a specific hardware pairing (the OPOTEK Phocus laser series and the Verasonics Vantage platform). This report does not represent the official specifications, endorsements, or views of OPOTEK, Inc. or Verasonics, Inc. The authors have made every effort to portray the technical behaviors of these systems accurately within the scope of our specific workflow. Should any inaccuracies or misrepresentations concerning the capabilities or features of OPOTEK or Verasonics products be identified, the authors are committed to addressing and correcting them promptly and in good faith.

Please note that the provided repository reconstructs the proposed elements in as general a form as possible, and therefore differs from the exact code utilized in our internal experiments. Furthermore, the sample code is provided ``as is'' without warranty of any kind. The authors do not guarantee its performance, executability, or suitability for any specific hardware or software environments.

\bibliographystyle{elsarticle-num}
\bibliography{report}

\end{document}